\documentclass{svproc}
\usepackage[utf8]{inputenc}
\usepackage{url}

\usepackage{natbib}
\bibpunct{(}{)}{;}{a}{}{,}%
\usepackage{graphicx}
\usepackage{amsmath,amssymb}
\usepackage{multirow}
\usepackage{longtable}
\newcommand{\msun}{\ensuremath{M_{\sun}}\,}

\usepackage[b5paper]{geometry}
\begin{document}
	\mainmatter              % start of a contribution
	\title{Confirmation of intermediate-mass black holes candidates with X-ray observations}
	\titlerunning{X-ray confirmation of intermediate-mass black holes candidates}  % abbreviated title (for running head)
	\author{Victoria Toptun\inst{1,3} \and Igor Chilingarian\inst{1,2} \and
		Kirill Grishin\inst{1} \and Ivan Katkov\inst{1,4} \and \\ Ivan Zolotukhin\inst{1} 
 \and Vladimir Goradzhanov\inst{1,3} \and
		Mariia Demianenko\inst{1,5} \and \\ Ivan Kuzmun\inst{3}}
	\authorrunning{Victoria Toptun et al.} % abbreviated author list (for running head)
	\tocauthor{Victoria Toptun, Igor Chilingarian, Kirill Grishin, Ivan Katkov, Ivan Zolotukhin, Vladimir Goradzhanov, Mariia Demianenko, and Ivan Kuzmun}
	\institute{Sternberg Astronomical Institute, Lomonosov Moscow State University, Universitetsky pr. 13, Moscow 119234, Russia\\
	\and Center for Astrophysics - Harvard and Smithsonian, 60 Garden Street MS09, Cambridge, MA 02138, USA \\
	\and Department of Physics, M.V. Lomonosov Moscow State University, 1 Vorobyovy Gory, Moscow 119991, Russia \\
	\and Center for Astro, Particle, and Planetary Physics, NYU Abu Dhabi, PO Box 129188 Abu Dhabi, UAE \\
	\and Moscow Institute of Physics and Technology, Institutskiy per. 9, Dolgoprudny 141701, Russia}
	
	\maketitle

%\begin{document}
%\renewcommand{\ConfSubsection}{Small and medium-size telescopes}
\begin{abstract}
The origin of supermassive black holes (SMBH) in galaxy centers still remains uncertain. There are two possible ways of their formation - from massive ($10^5 - 10^6$ \msun) and low-mass (100 \msun) BH nuclei. The latter scenario should leave behind a large number of intermediate mass black holes (IMBH, $10^2 - 10^5$ \msun). The largest published sample of bona-fide IMBH-powered AGN contains 10 objects confirmed in X-ray. Here we present a new sample of 15 bona-fide IMBHs, obtained by confirming the optically selected IMBH candidates by the presence of radiation from the galactic nucleus in the X-ray range, which increases the number of confirmed IMBHs at the centers of galaxies by 2.5 times. In the same way, 99 black holes with masses of $2\cdot10^5 - 10^6$ \msun were confirmed. The sources of X-ray data were publicly available catalogs, archives of data, and our own observations on XMM-Newton, Chandra and Swift. The Eddington coefficients for 30\% of the objects from both samples turned out to be close to critical, from 0.5 to 1, which is an unusually high fraction. Also for the first time for light-weight SMBH the correlations between the luminosity in the [OIII] emission line or the broad component of the H$\alpha$ line and the luminosity in the X-ray range were plotted.
	% We would like to encourage you to list your keywords within
	% the abstract section using the \keywords{...} command.
\end{abstract}

\section{Introduction}
% TODO: переписать своими словами и исправить ошибки
Massive black holes in galaxy centres are believed to co-evolve with the spheroids of their hosts \citep{kormendy13}, growing via coalescences during galaxy mergers \citep{2005LRR.....8....8M} and by gas accretion \citep{2012Sci...337..544V}. The discovery of quasars in the early Universe (z $>6.3$, only 750-900 Myr after the Big Bang) hosting super-massive black holes (SMBHs) as heavy as $10^{10}$\msun \citep{mortlock11,2015Natur.518..512W} cannot be explained by gas accretion onto stellar mass black hole seeds ($<100$\msun) alone. Alternatively, the rapid inflow and subsequent direct collapse of gas clouds \citep{loeb94,begelman06}  can form massive seeds (M $>10^5-10^6$\msun). The latter scenario solves the SMBH early formation puzzle butleads to a gap in the present-day black hole mass function in the IMBH regime ($100 < M_{BH} <2\cdot10^5$\msun), whereas stellar mass seeds should leave behind a large number of IMBHs. Therefore, the elusive IMBH population holds a clue to the understanding of SMBH formation. Now the largest published sample of bona-fide IMBH-powered AGN contains 10 objects confirmed in X-ray.
% написать про то что уже были кандидаты в imbh и подтвержденные imbh но их мало

\section{Sample description}
% копипаст+транслейт курсача; исправить
The work is based on a sample of 1928 galaxies with active nuclei with virial masses of black holes up to $10^6$\msun, 305 of which are IMBHs with masses less than $2\cdot10^5$ \msun. The sample was obtained as part of the work on the article \citep{Chilingarian+18} by analyzing the spectra of about a million galaxies from the SDSS DR7 \citep{SDSS_DR7} survey. A broad component of the $H_{\alpha}$ emission line was identified in the spectra, and the virial masses of the central black holes were calculated from it's parameters. A method for selecting BHs of intermediate masses and light-weight SMBH was proposed in the work \citep{Chilingarian+18}. Also only galaxies with well-measured central BH masses and reliable broad component in lines were selected, which ratio of the BH mass to the mass error is more than 3, important lines do not overlay on the sky lines due to the redshift, and falls into the AGN or transition region on the BPT \citep{BPT81} diagnostic diagram.

\section{Data analysis methods}
\subsection{X-ray data collection, calibration and analysis}
To confirm the AGN nature of an object it is necessary to detect the X-ray emission, produced by the accretion disk of the black hole. X-ray data were obtained from the Chandra Source Catalog 2, 4XMM-DR10, Second Swift XRT Point Source Catalog, Second ROSAT all-sky survey, from the x-ray data archives as Chandra Data Archive and XMM-Newton science archive and our own observations at XMM-Newton, Chandra and Swift. For calibration of raw observational data and extraction of spectra was used special software: Science Analysis System (SAS) for XMM-Newton data, Chandra Interactive Analysis of Observations for Chandra and XSelect for Swift. 

The next important step was extracting the spectrum of the object and fitting its parameters. Using the parameters and form of the spectrum we can once again confirm its AGN nature. Spectral data was fitted with powerlaw and photoelectric absorption models with XSPEC and Sherpa packages in Python 3. It turned out that most of our objects have a photon index over 2, that is typical for AGN.
\vspace{-0.5cm}
\begin{figure}
\includegraphics[width=0.5\hsize]{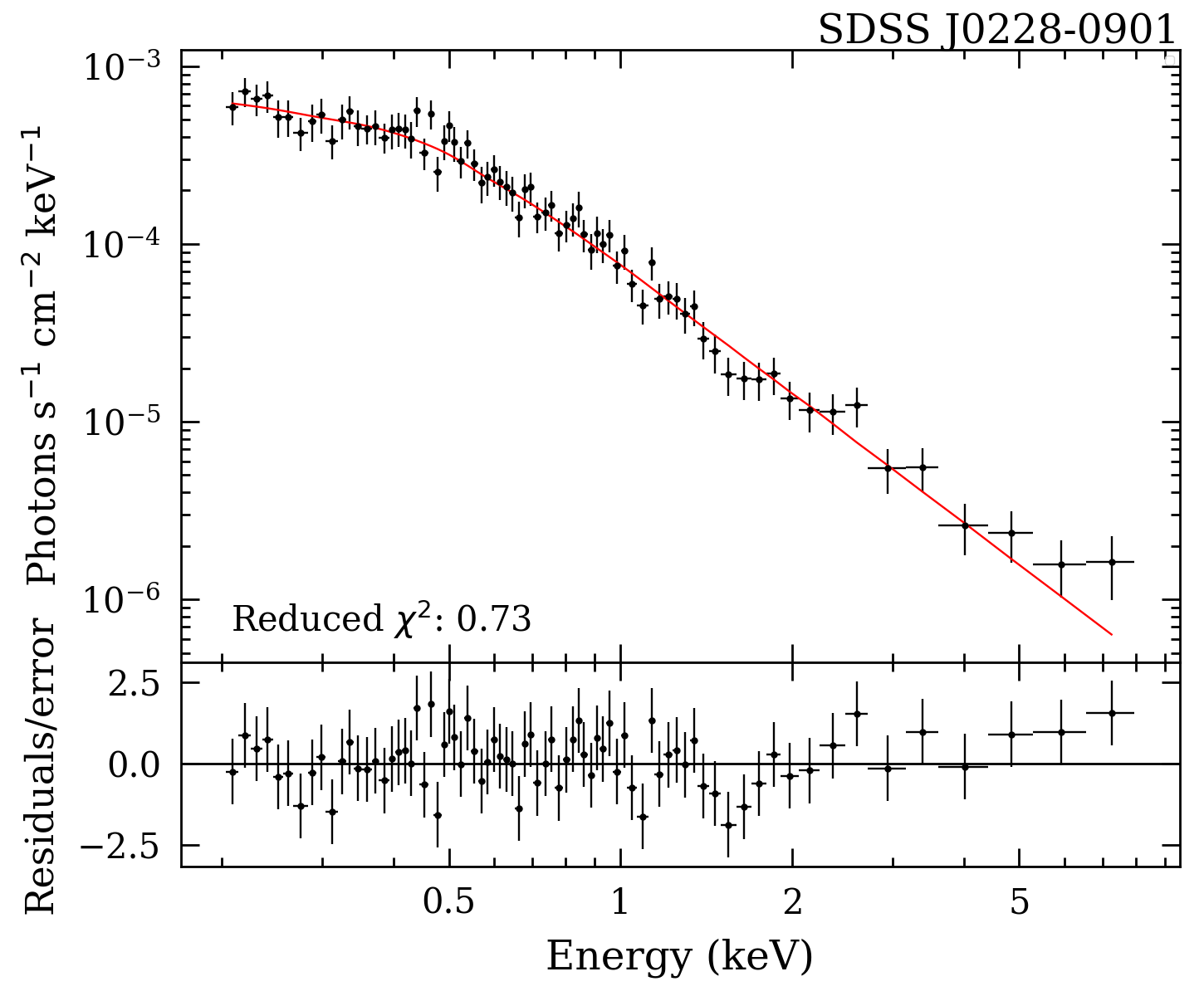} 
\includegraphics[width=0.5\hsize]{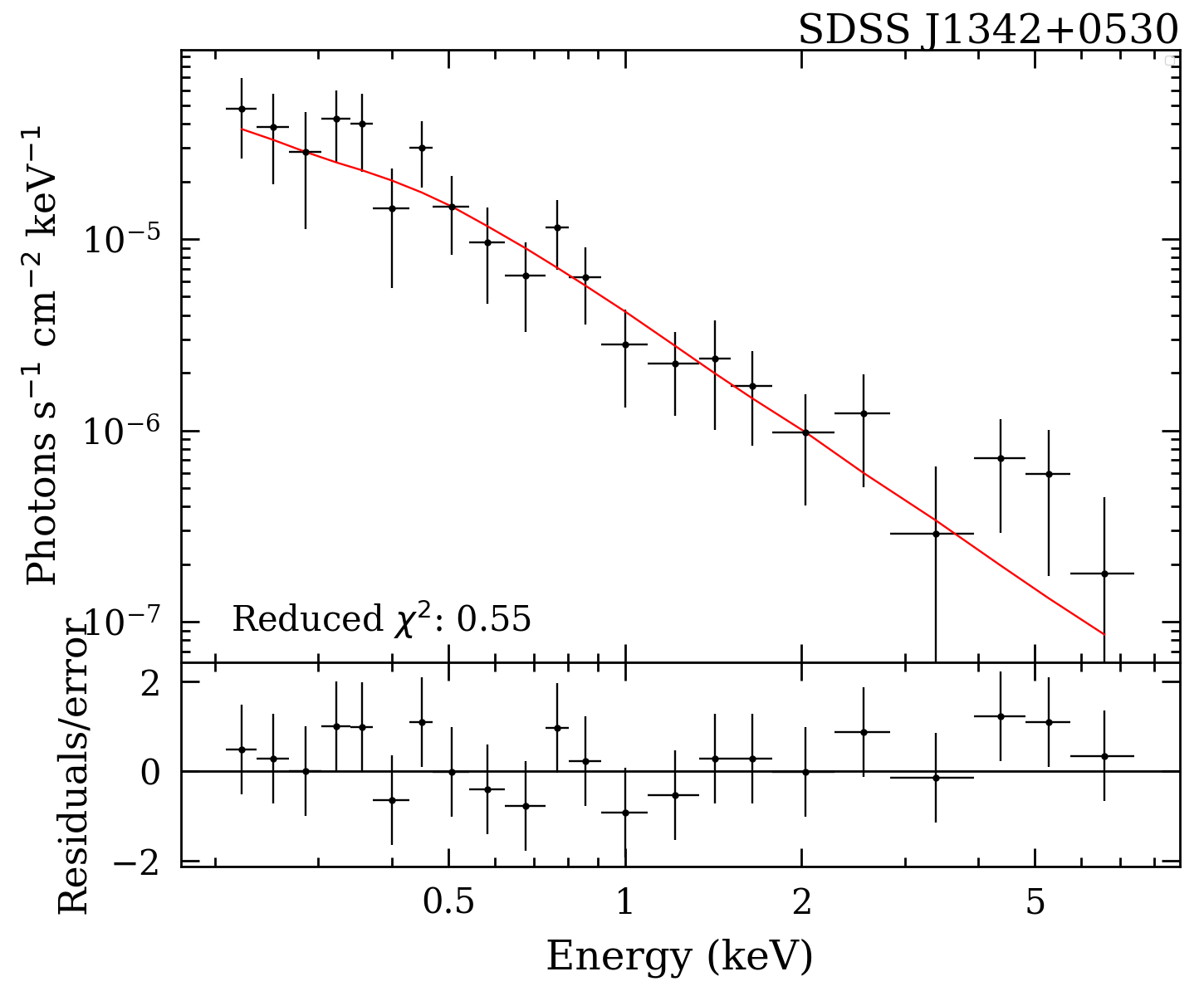} \\
\vspace{-0.5cm}
\caption{Examples of spectra. The model is shown with a red line
\label{spec}}
\vspace{-1cm}
\end{figure}

\subsection{X-ray luminosity of the stellar population}
It is worth remembering that the object’s X-ray luminosity should be higher than expected from the stellar population of the host galaxy. X-ray stellar population mainly consists of low-mass and high-mass X-ray binaries. Their contribution to luminosity can be calculated using the following relations
$$L_{HX}^{gal} = \alpha M_*+\beta SFR$$
$$\alpha = (9.05\pm0.37)\times10^{28}\,ergs\,s^{-1}\,M_{\odot}^{-1}$$
$$\beta = (1.62\pm0.22)\times10^{39}\,ergs\,s^{-1}\,(M_{\odot}\,yr^{-1})^{-1}$$

\subsection{Eddington luminosities}
The luminosity of the accretion disk directly depends on this accretion rate. Therefore, in order to understand how high the rate of accretion onto a black hole is, it is necessary to calculate the Eddington ratio. The masses are known from optical observations and the Eddington luminosity is easily calculated
$$L_{edd} = \frac{4\pi GMm_pc}{\sigma_T}\approx 10^{38}\frac{M}{M_{\odot}}\:ergs\cdot s^{-1}$$

The bolometric luminosity values are determined from the luminosity in the X-ray range through the bolometric correction coefficient defined as
$$K(L_X) = \frac{L_{bol}}{L_X} = 15.33\left[ 1 + \left( \frac{\log(L_X/L_{\odot}}{11.48} \right)^{16.20}\, \right]$$

\section{Results and conclusions}
\subsection{New IMBHs with X-ray confirmation}
From the sample of 1928 candidates for low-mass black holes, 124 black holes with masses less than $10^6\,$\msun were confirmed in the X-ray range, of which 24 are intermediate masses black holes with masses $M_{BH}<2\cdot10^5 \,$\msun, and their x-ray luminosities were calculated. For 55 objects from this sample, in addition to the luminosity, the parameters of their X-ray spectra were determined. Thus, confirmed intermediate-mass black holes sample was increased by a factor of 2.5, and this number of IMBHs is a strong argument for the formation of supermassive black holes from low-mass progenitors. 
Most of the sources have ''soft'' X-ray spectra -- for 39 of 55 objects the photon index is $\Gamma>2$. For a large number of objects the Eddington  ratio turned out to be very high. Thus, for 39 of 123 BHs it is more than 30\%. This points on very high accretion rates on many IMBH.

\subsection{Correlation between X-ray luminosity and luminosity in emission lines}
For supermassive black holes have found a correlation between the luminosity in the X-ray range and the optical emission lines [OIII] and $H_{\alpha}$ \citep{2005ApJ...634..161H, 2015ApJ...815....1U, 2012MNRAS.422.3268J}. We verified this correlation for light-weight SMBH and IMBH for the first time. The fluxes in lines were obtained together with the black holes masses by processing optical spectra. Correlations were approximated by linear regression $$\log(L_{emis.\:Line})=a\cdot\log(L_X)+b$$ The results of this approximation are shown in the table \ref{regression}.
\begin{table}[h!]
\vspace{-0.5cm}
\caption{Aproximation results\label{regression}}
\centering
\begin{tabular}{ | c | c | c | c | }
\hline
Line & a & b & error \\ \hline
[OIII] & $0.339\pm0.045$ & $25.950\pm1.898$ & 0.475 \\
$H_{\alpha}$ & $0.344\pm0.039$ & $25.680\pm1.642$ & 0.411 \\
\hline
\end{tabular}
\vspace{-0.5cm}
\end{table}

It turned out that for low-mass black holes the slope of the regression is much smaller than for the sample of supermassive black holes. It could be caused by a change in the geometry of the region of X-ray radiation - this and other possible hypotheses requires further research. However, it is already clear that these correlations are extremely useful - it can be used to estimate the X-ray luminosity for calculating the exposures necessary to accumulate a sufficient number of photons for spectral analysis.
\vspace{-0.5cm}
\begin{figure}
\includegraphics[width=0.5\hsize]{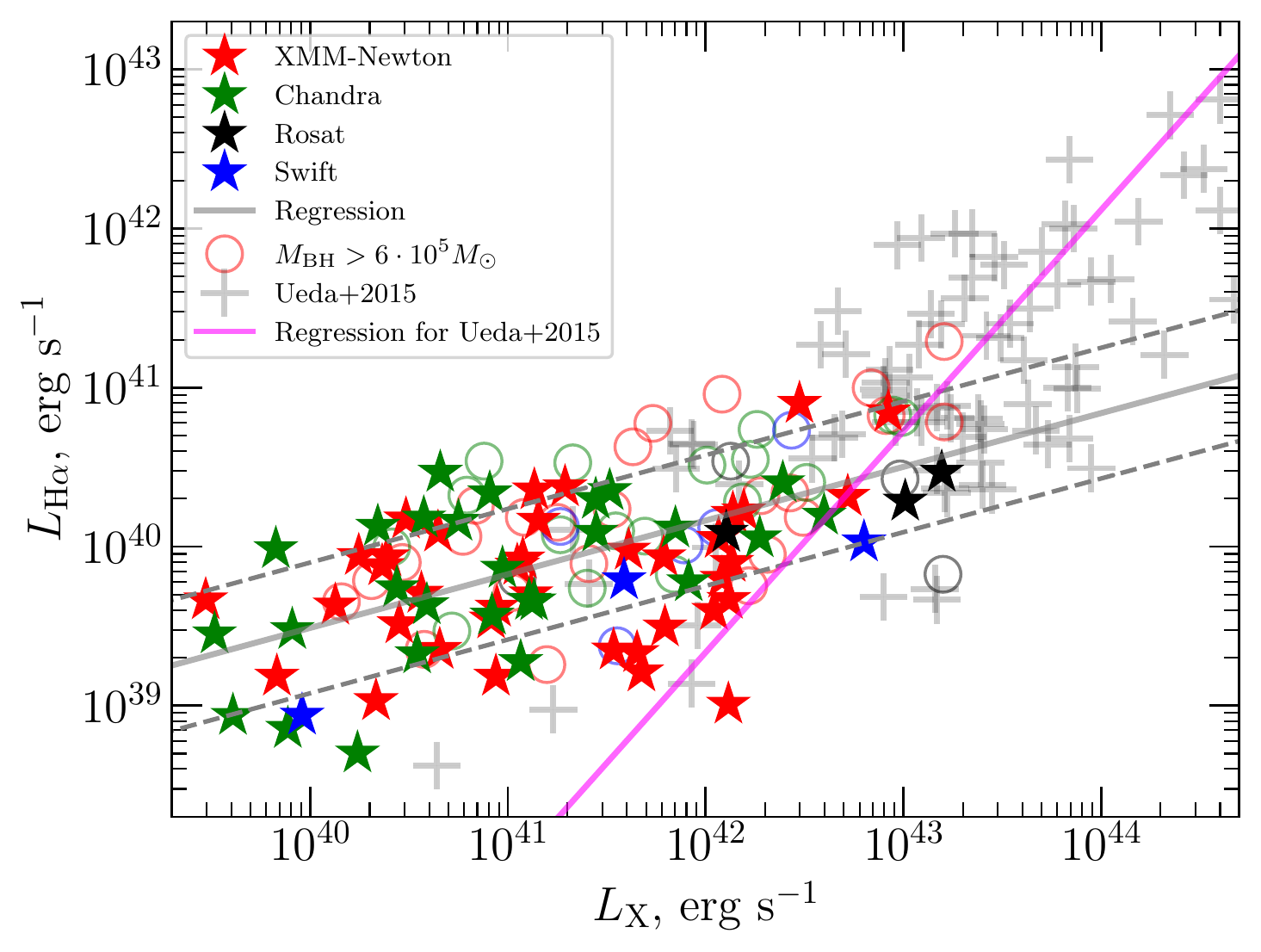}
\includegraphics[width=0.5\hsize]{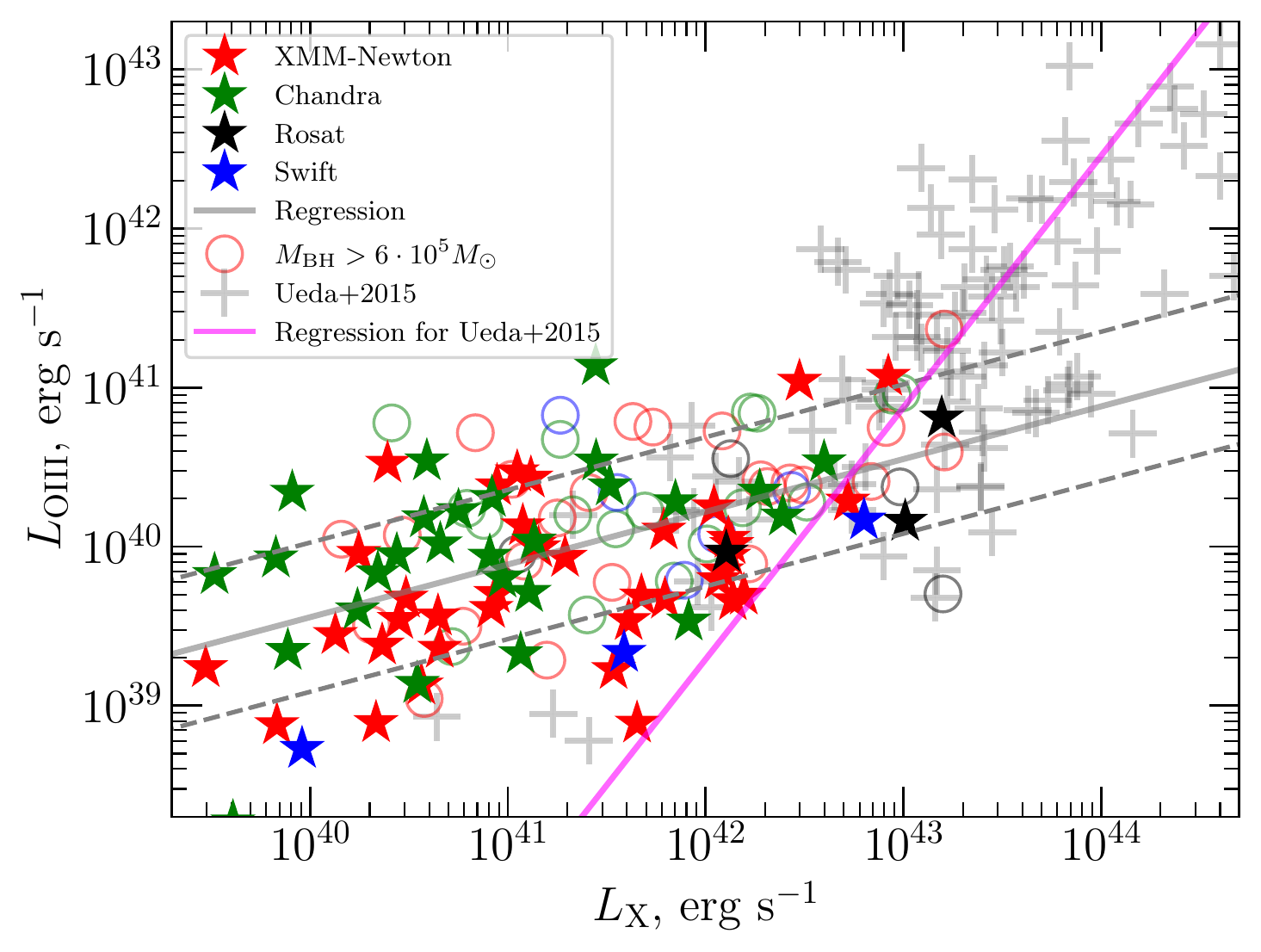} \\
\vspace{-0.5cm}
\caption{Correlations between the luminosity in the [OIII] emission line or the broad component of the H$\alpha$ line and the luminosity in the X-ray range
\label{figsample}}
\vspace{-1cm}
\end{figure}

\subsection{Conclusions about the ways of black holes formation}

The key result of this study is a significant increase in sample of IMBH, and this confirms that SMBH grew from their lower mass progenitors. Also we show that at low mass regime the number of objects emitting close to the Eddington limit is unexpectedly large. This indicates that BH can grow rapidly, including due to accretion of matter, not only due to merging and host galaxies bulges growth.

\textbf{Acknowledgements.} This project is supported by the RScF Grant 17-72-20119.

%\begin{thebibliography}{6}
%	%
%	
%	\bibitem {smit:wat}
%	Smith, T.F., Waterman, M.S.: Identification of common molecular subsequences.
%	J. Mol. Biol. 147, 195?197 (1981). \url{doi:10.1016/0022-2836(81)90087-5}
%	
%	\bibitem {may:ehr:stein}
%	May, P., Ehrlich, H.-C., Steinke, T.: ZIB structure prediction pipeline:
%	composing a complex biological workflow through web services.
%	In: Nagel, W.E., Walter, W.V., Lehner, W. (eds.) Euro-Par 2006.
%	LNCS, vol. 4128, pp. 1148?1158. Springer, Heidelberg (2006).
%	\url{doi:10.1007/11823285_121}
%	
%	\bibitem {fost:kes}
%	Foster, I., Kesselman, C.: The Grid: Blueprint for a New Computing Infrastructure.
%	Morgan Kaufmann, San Francisco (1999)
%	
%	\bibitem {czaj:fitz}
%	Czajkowski, K., Fitzgerald, S., Foster, I., Kesselman, C.: Grid information services
%	for distributed resource sharing. In: 10th IEEE International Symposium
%	on High Performance Distributed Computing, pp. 181?184. IEEE Press, New York (2001).
%	\url{doi: 10.1109/HPDC.2001.945188}
%	
%	\bibitem {fo:kes:nic:tue}
%	Foster, I., Kesselman, C., Nick, J., Tuecke, S.: The physiology of the grid: an open grid services architecture for distributed systems integration. Technical report, Global Grid
%	Forum (2002)
%	
%	\bibitem {onlyurl}
%	National Center for Biotechnology Information. \url{http://www.ncbi.nlm.nih.gov}
%	
%	
%\end{thebibliography}

\bibliographystyle{main}
\bibliography{main}
\end{document}